\documentstyle[twocolumn,prb,aps,epsf]{revtex}
\begin{document}
\draft
\twocolumn[\hsize\textwidth\columnwidth\hsize\csname @twocolumnfalse\endcsname
%
%
%

\title{Sub-Dominant Pairing Channels in Unconventional Superconductors: 
Ginzburg-Landau Theory }

\author{Qiang-Hua Wang}

\address{Department of Physics and National Laboratory of Solid State Microstructures,\\
Institute for Solid State Physics, Nanjing University, Nanjing 210093, China\\
Department of Physics, University of Hong Kong, Pokfulam Road, Hong Kong, China}
\author {Z. D. Wang$^*$ and Q. Li}
\address{Department of Physics, University of Hong Kong, Pokfulam Road, Hong Kong, China}

\maketitle

\begin{abstract}
A Ginzburg-Landau theory is developed for unconventional superconductors
with the three
relevant singlet pairing channels (i.e., $s$-, $d_{x^2-y^2}$- and $d_{xy}$
channels). Various consequences of the sub-dominant channels (i.e., $s$- and
$d_{xy}$-channels) are examined in detail. (1) In the case of a
$d_{x^2-y^2}+is$-wave superconductor, we reproduce an earlier result that
there is a second order zero-field transition from the pure $d_{x^2-y^2}$-phase to the
time-reversal-symmetry- ($\top$-) breaking $d_{x^2-y^2}+is$-phase at the
temperature $T_{DS}$. The structure of a single vortex above and below $T_{DS}$
is four-fold and two-fold symmetric, respectively. (2) In the case
of a $d_{x^2-y^2}+id_{xy}$-wave superconductor, there is also a second
order zero-field phase transition from the pure $d_{x^2-y^2}$-phase to the
$\top$-breaking $d_{x^2-y^2}+id_{xy}$-wave phase at the temperature $T_{DD'}$.
 In contrast to the case in a $d_{x^2-y^2}+is$-wave
superconductor, the subdominant phase can {\it not} be induced by vortices
above $T_{DD'}$. Below the $\top$-breaking
transition, the sub-dominant phase in the mixed state is nontrivial: it
survives at low fields, but may disappear above a field (increasing with
decreasing temperature) presumably via a first-order transition. 
(3)By including the strong coupling effects,
 a $\top$-breaking coupling term between
the $d_{x^2-y^2}$- and $d_{xy}$-waves is  found to have significant effects on
the low temperature behavior of $d_{x^2-y^2}+id_{xy}$ superconductors.
In a magnetic field, a $d_{x^2-y^2}+id_{xy}$
state is always established, but the field-dependence of 
$d_{xy}$-amplitude above $T_{DD'}$ is different from that below $T_{DD'}$.
Above but not very close to $T_{DD'}$, the induced minimum gap $\Delta_0\propto
B/(T-T_{DD'})$.
\end{abstract}

]
\vskip10pc
\narrowtext
\section{Introduction}
The phase-sensitive experiment of the unusual Josephson effect
in high-$T_c$ superconductors evidenced the $d_{x^2-y^2}$-wave
symmetry of Cooper pair wave functions.~\cite{Tsui}
Thus at least the problem as
what is the dominant channel for electrons to pair in high-$T_c$
superconductors has been resolved.
However, a new puzzle arises as what is the sub-dominant pairing channel,
which seems to have manifested in a number of experiments, e.g., the
observation of surface-induced broken time-reversal-symmetry
($\top$ hereafter) in YBCO tunnel junctions,\cite{Covington-Fogelstrom}
the observation of fractional vortices trapped
in a boundary junction,\cite{Kirtly}
and the abnormal field dependence of the low temperature
thermal conductivity $\kappa_e$ in BSCCO,~\cite{Krishana,Aubin} to name
only a few. At surfaces, the sub-dominant channel can show up because of
a suppression of the effective critical temperature in the
dominant $d_{x^2-y^2}$-channel. Rather differently,
the anomalous $\kappa_e$ reported in Ref.\cite{Krishana}
might require a mechanism to produce {\it bulk} $\top$-breaking states.
~\cite{Laughlin,Ghosh}
This scenario might
work above $5K$ in the samples reported in Ref.~\cite{Krishana,Aubin}
but failed to explain the data at sub-Kelvin temperatures in Ref.~\cite{Aubin}.
Instead, field-induced quasiparticle excitations with pure $d_{x^2-y^2}$-wave
gap function explain qualitatively these data.~\cite{Aubin}

As a modeling study, we consider the relevant {\it singlet} sub-dominant
pairing channels in an unconventional $d$-
superconductor to be the $s$- and $d_{xy}$-channels, which may be relevant
to high temperature superconductors. For attractive pairing interactions
in the $s$- and $d_{x^2-y^2}$-channels, or in the $d_{x^2-y^2}$- and
$d_{xy}$-channels, the the uniform superconducting state
in the absence of magnetic field has been shown to be either a pure
state in the dominant channel, or a state with a broken
time-reversal-symmetry involving the relevant channels.~\cite{Ghosh,Joynt}
In order to study the
vortex states, we develop a Ginzburg-Landau (GL) theory that includes
the above-mentioned sub-dominant channels. We discuss the consequences of
these channels on the properties of the superconductors, such as the
field dependence of the thermal conductivity.
The structure of this paper is as
follows. The microscopic derivation of the GL theory (in the weak coupling
limit) is presented in section II. The properties of $d_{x^2-y^2}+is$
superconductors are described in section III. The properties of $d_{x^2-y^2}
+id_{xy}$ superconductors are addressed in section IV. A phenomenological
strong coupling correction to the weak coupling theory is developed and
discussed in section V with respect to the $d_{x^2-y^2}+id_{xy}$
superconductors. Section VI contains a summary and some concluding remarks.

\section{Derivation of the Ginzburg-Landau theory in the weak coupling limit}
The simplest way to derive microscopically a Gingzburg-Landau theory
for superconductors is to consider the Bardeen-Cooper-Shrieffer (BCS)
gap equation. At zero center-of-mass momentum for the Cooper pairs,
the gap function is
\begin{eqnarray}
\Delta_k=\sum_{k'}V_{k,k'}\langle C_{-k',\downarrow}C_{k',\uparrow}\rangle
\nonumber\\
=\sum_{k'}\frac{V_{k,k'}\Delta_{k'}}{2E_{k'}}\tanh\frac{\beta E_{k'}}{2}
\label{Eq:Gap}
\end{eqnarray}
which follows routinely from the effective BCS Hamiltonian.
Here $E_k=\sqrt{\epsilon_k^2+|\Delta_k|^2}$ is the quasi-particle excitation
spectrum, and $\beta=1/T$ is the inverse temperature.
(We are using the units $k_B=\hbar=c=1$, and are measuring energies
from the Fermi surface.) In the weak coupling
limit the summation over momentum in the gap equation is understood to be
restricted near the Fermi surface. Before going into further details, we
need to specify the pairing interaction $V_{k,k'}$.
Due to the tetragonal symmetry of
the copper planes, it is rather reasonable to assume the following form for
$V_{k,k'}$,
\begin{eqnarray}
V_{k,k'}&=&V_s+V_{x^2-y^2}(\hat{k}_x^2-\hat{k}_y^2)(\hat{k'}_{x}^2-\hat{k'}_{y}^2)
\nonumber\\&+&4V_{xy}\hat{k}_x\hat{k}_y\hat{k'}_{x}\hat{k'}_{y}\nonumber\\
&=&V_s+V_{x^2-y^2}\cos 2\theta_k\cos 2\theta_{k'}\nonumber\\
&+&V_{xy}\sin 2\theta_k
\sin 2\theta_{k'}.\label{Eq:Vkk'}
\end{eqnarray}
Henceforth the subscripts $_s$, $_{x^2-y^2}$ and $_{xy}$ are channel indices
referring to the $s$-, $d_{x^2-y^2}$- and $d_{xy}$-channels of the pairing
interaction, respectively. $\theta_k$ is the angle between the wave vector
$k$ and the $a$-axis in the copper-oxide plane. By symmetry, the gap function can be expressed as
\begin{eqnarray}
\Delta_k=S+D\cos 2\theta_k+D'\sin 2\theta_k,\label{Eq:Dk}
\end{eqnarray}
where $S$, $D$ and $D'$ are complex amplitudes of the
gap function in the $s$-, $d_{x^2-y^2}$- and $d_{xy}$-channels,
respectively, and are also called as the {\it order
parameters}. Inserting Eqs.~(\ref{Eq:Vkk'}) and (\ref{Eq:Dk})
into Eq.~(\ref{Eq:Gap}), expanding the gap equation to the third order
in the order parameters [using the Poisson identity $(1/2E)\tanh(\beta E/2)\equiv T\sum_n (\omega_n^2+E^2)^{-1}$
with $\omega_n$ being the Fermion
Matsubara frequency], we obtain the GL equations in the uniform case,
\begin{eqnarray}
0&=&-N(0)V_s\left\{2\alpha_SS +\gamma\left[2|S|^2 S+|D|^2 S +|D'|^2 S\right.\right.\nonumber\\
&+&\left.\left.(S^*D+{\rm c.c})D+(S^*D' +{\rm c.c.})D'\right]\right\};\nonumber\\
0&=&-N(0)V_{x^2-y^2}\left\{\alpha_{D}D+\gamma\left[\frac{3}{4}|D|^2 D+\frac{1}{4}|D'|^2 D\right.\right.\nonumber\\
&+&|S|^2 D+\left.\left.(S^* D+{\rm c.c.})S+\frac{1}{4}(D^*D'+{\rm c.c.})D'\right]\right\};\label{Eq:homo}\\
0&=&-N(0)V_{xy}\left\{\alpha_{D'}D'+\gamma\left[\frac{3}{4}|D'|^2 D'+\frac{1}{4}|D|^2 D'\right.\right.\nonumber\\
&+&|S|^2 D'+\left.\left.(S^* D'+{\rm c.c.})S+\frac{1}{4}(D^*D'+{\rm c.c.})D\right]\right\}.\nonumber
\end{eqnarray}
Here $N(0)$ is the normal state density of states (DOS) at the Fermi surface, $\alpha_i=\ln T/T_i$ ($i=S,D,D'$) with $T_{i}$ the bare critical
temperatures at the $i$-channel: $T_{S}\approx 
1.14\omega_c\exp(-1/N(0)V_s)$,
and $T_{D,D'}\approx 1.14\omega_c\exp(-2/N(0)V_{x^2-y^2,\,xy})$ (
with $\omega_c$ the energy cut-off for the pairing interaction).
Finally $\gamma=\int_{-\omega_c}^{\omega_c}d\epsilon T\sum_n (\omega_n^2+\epsilon^2)^{-2}$.
On the other hand, the rigidity of the order parameters can
be probed by the long wave-length ($\gg\lambda_F$)
spatial variation of the order parameters.
For this purpose, it suffices to consider
the linearized gap equation but with a small center-of-mass momentum $q$
($\ll k_F$) for each Cooper pair,
\begin{eqnarray*}
\Delta_{k,q}=\sum_{k'}\frac{V_{k,k'}\Delta_{k',q}}{\epsilon_{k'_+}
+\epsilon_{k'_-}}\left[1-f(\epsilon_{k'_+})-f(\epsilon_{k'_-})\right],
\end{eqnarray*}
where $k_{\pm}=k\pm q/2$, and $f(\epsilon)$
is the Fermi distribution function.
We now assume $\Delta_{k,q}=S_q+D_q\cos 2\theta_k+D'_q\sin 2\theta_k$, and
expand the above equation to the second
order in $q$ to find that the Fourier components $S_q$, $D_q$ and $D'_q$ satisfy
\begin{eqnarray*}
0&=&-N(0)V_s\left\{2\alpha_S S_q+K\left[ 2q^2 S_q+(q_x^2-q_y^2)D_q
\right.\right.\\
&+&\left.\left.(q_xq_y+q_yq_x)D'_q\right]\right\};\\
0&=&-N(0)V_{x^2-y^2}\left\{\alpha_D D_q+K\left[ q^2 D_q+(q_x^2-q_y^2)S_q
\right]\right\};\\
0&=&-N(0)V_{xy}\left\{\alpha_{D'}D'_q+K\left[q^2 D'_q+(q_xq_y+q_yq_x)S_q
\right]\right\},
\end{eqnarray*}
where $K=v_F^2\gamma/8$ reflects the rigidity of the order parameters
with $v_F$ being the Fermi velocity.
Resorting to real space, and using the gauge invariant gradient operator
$\Pi=-i\nabla-2eA$ (with $A$ the vector potential)
in substitution of $-i\nabla$,
and finally combining the nonlinear homogeneous tems in
Eq.(\ref{Eq:homo}), we obtain the complete GL equations (up to
third order in the order parameters) describing pairing in all of
the three most relevant spin-singlet channels,
\begin{eqnarray}
& &2\alpha_S S+\gamma\left[ 2|S|^2S+|D|^2S+|D'|^2S+(S^*D+{\rm c.c.})D\right.\nonumber\\
& &\left.+(S^*D'+{\rm c.c.})D'\right] +K\left[2\Pi^2 S+(\Pi_x^2-\Pi_y^2)D\right.\nonumber\\
& &+\left.(\Pi_x\Pi_y+\Pi_y\Pi_x)D'\right]=0;\nonumber\\
& &\alpha_D D +\gamma\left[ \frac{3}{4}|D|^2 D+\frac{1}{4}|D'|^2D+|S|^2D\right.\nonumber\\
& &\left.+(S^*D+{\rm c.c.})S+\frac{1}{4}(D^*D'+{\rm c.c.})D'\right]\nonumber\\
& &+K\left[\Pi^2 D+(\Pi_x^2-\Pi_y^2)S\right]=0;\label{Eq:GL}\\
& &\alpha_D' D'+\gamma\left[\frac{3}{4}|D'|^2D'+\frac{1}{4}|D|^2D'+|S|^2 D'\right.\nonumber\\
& &\left.+(S^*D'+{\rm c.c.})S+(D^*D'+{\rm c.c.})D\right]\nonumber\\
& &+K\left[\Pi^2 D'+(\Pi_x\Pi_y+\Pi_y\Pi_x)S\right]=0. \nonumber
\end{eqnarray}
For completeness and general purposes, we need a GL free energy
functional. Following from Eqs.~(\ref{Eq:GL}), it is given by~\cite{Note}
\begin{eqnarray}
F=& &\frac{N(0)}{2}\int_{\Omega} 2\alpha_S|S|^2+\alpha_D|D|^2+\alpha_{D'}|D'|^2\nonumber\\
& &+\gamma\left[ |S|^4+\frac{3}{8}|D|^4+\frac{3}{8}|D'|^4+|S|^2|D|^2\right.\nonumber\\
& &+|S|^2|D'|^2+\frac{1}{4}|D|^2|D'|^2+\frac{1}{2}(S^*D+{\rm c.c.})^2\nonumber\\
& &\left.+\frac{1}{2}(S^*D'+{\rm c.c.})^2+\frac{1}{8}(D^*D'+{\rm c.c.})^2\right]\nonumber\\
& &+K(2|\Pi S|^2+|\Pi D|^2+|\Pi D'|^2)\nonumber\\
& &+K\left\{(\Pi_x\hat{x}+i\Pi_y\hat{y})D\cdot
\left[(\Pi_x\hat{x}-i\Pi_y\hat{y})S\right]^*
+{\rm c.c.}\right\}\nonumber\\
& &+K\left\{
(\frac{\Pi_x+\Pi_y}{\sqrt{2}}\hat{x}+i\frac{\Pi_x-\Pi_y}{\sqrt{2}}\hat{y})D'\right.\nonumber\\
& &\left.\cdot[\frac{\Pi_x+\Pi_y}{\sqrt{2}}\hat{x}
-i\frac{\Pi_x-\Pi_y}{\sqrt{2}}\hat{y})S]^* +{\rm c.c.}\right\}\nonumber\\
& &+\int_{\Omega} (\nabla\times A)^2/8\pi \label{Eq:Fenergy},
\end{eqnarray}
where uniformity along the $c$-axis is assumed, and
$\int_{\Omega}$ denotes integration over the $ab$-plane.
The prefactor of $N(0)/2$ could be obtained from a microscopic
derivation of the super-current, but it can also be obtained by
the inspection that the free energy reduces to the superconducting
ground state energy at zero temperature. The last term in $F$ is the
magnetic energy. The equilibrium super-current can now be easily obtained
from the functional derivative
of the free energy with respect to the vector potential,
$\delta F/\delta A=0$. It can be seen clearly that the GL equations
(or the free energy functional)
would be identical to that derived previously for $d_{x^2-y^2}+is$
superconductors if the $d_{xy}$-channel were shut-down,
and would be equivalent to that investigated for $d_{x^2-y^2}+id_{xy}$
superconductors if the $s$-wave channel were disregarded.
Moreover, the gradient terms for $D$ and $D'$ are symmetric with respect
to each other under a rotation of $\pi/4$.

\section{$d_{x^2-y^2}+is$ superconductors}
This case has been studied previously,~\cite{Ren-Xu,Berlinsky,Franz,Wang}
so that we shall first recapitulate some essential points that interest us,
and add some novel discussions.

We assume that $T_D>T_S$ as is the case in high-$T_c$
superconductors. Define $\alpha_{SD}=\alpha_S/|\alpha_D|$
with $\alpha_D<0$. Optimizing the free energy Eq.~(\ref{Eq:Fenergy})
one finds that the uniform bulk phase
below the highest bare critical temperature $T_D$
is a pure $D$ phase at $\alpha_{SD}>-2/3$, or above a critical
temperature $T_{SD}=T_S^3/T_D^2$,
at which the system undergoes a second-order phase transition to a
$\top$-broken phase $(S,D)\propto
(\pm i\eta,1)$, with $\eta=|S|/|D|=
\sqrt{-(3\alpha_{SD}+2)/(4+2\alpha_{SD})}$.

Although there is no uniform $S$ at $T>T_{SD}$,
$S$ can be induced by inhomogeneities, such as surfaces (or twinboundaries),
disorders and vortices, due to the mixed gradient terms in
Eq.~(\ref{Eq:Fenergy}). Of special interest is the $S$-component generated
by vortices. At $T>T_{SD}$, the $S$-component is sub-dominant and
vanishes where $D$ is uniform (e.g., far from the vortex core). Thus it can
be obtained perturbatively. Assume $D\sim e^{i\theta}$ in cylindrical
coordinates and in the gauge $A=A(r)\hat{\theta}$. Since
$\Pi_x^2-\Pi_y^2$ transforms as $e^{\pm 2i\theta}$
and scales as $1/r^2$ at $r\rightarrow\infty$, we immediately see that
to the first order $S\sim a(r)e^{3i\theta}+b(r)e^{-i\theta}$ (the concrete
forms of $a(r)$ and $b(r)$ are unimportant, except that both $a(r)$
and $b(r)$ vanishes as $r$ and $1/r^2$ as $r\rightarrow 0$ and $\infty$,
respectively) so that $|S|^2\sim a(r)^2+b(r)^2+2a(r)b(r)\cos 4\theta$.
(For complex $a$ and $b$, there are $\theta$-independent phase shifts in
the argument of the cosine function, but they
do not alter our general conclusion. This is also the case in similar
discussions hereafter). Therefore $|S|^2$ is
four-fold symmetric near the vortex core and vanishes far from the
core. This result has been obtained previously by many authors,
~\cite{Ren-Xu,Berlinsky,Franz,Volovik} in different contexts, and has been
verified by our numerical simulations.~\cite{Li,Wang} In particular,
Franz {\it et al} were able to obtain explicit analytical as well as
numerical solutions for the single vortex state,~\cite{Franz}
which is useful for a quantitative comparison between the theory and the 
experiments. The relative phase of these two components varies continuously
around the vortex core. However, such a kind of vortex-induced $S$
is insufficient, at least at low fields, to change the bulk quasi-particle
DOS $N(\omega)\sim\frac{1}{\Omega}
\int_{\Omega}\sum_k\delta(\omega-\sqrt{\epsilon_k^2+|\Delta_k|^2})$
in that the induced $S$-component is localized near the vortex core.
Moreover, the behavior of $S$ does not show any sign of {\it increasing
sensitivity} to the applied field at decreasing temperatures.
Therefore, we may rule out the role of the $s$-channel in the
abnormal thermal conductivity of Krishana {\it et al}.~\cite{Krishana}

On the other hand, at $T<T_{SD}$, $\top$ is already broken in the
uniform phase. Lowest order perturbative treatment of $S$ is no longer valid
at $T\ll T_{SD}$. $S$ and $D$ are now of competing order far from
the core with a locked relative phase $\pm\pi/2$ because
of the fact that they are uniform there. In other words,
symmetry consistency requires that $S$ develops a winding component
$\sim e^{i\theta}$ in addition to the $e^{-i\theta}$- and $e^{3i\theta}$-
components, i.e., $S\sim ae^{i\theta}+be^{3i\theta}
+ce^{-i\theta}+$higher order terms near the vortex core. Thus generally
$|S|^2=a^2+b^2+c^2+2(ab+ac)\cos 2\theta+2bc\cos 4\theta$, developing
a striking two-fold, rather than a four-fold,
symmetry near the vortex core. The situation is similar for the $D$-component.
The relative phase varies again near and around the vortex, but is locked to
$\pm\pi/2$ far away from the core. As the two-fold symmetry is generated
solely from $ae^{i\theta}$, the quantity $a(r\rightarrow\infty)=\pm i\eta$
can also be thought of as the order parameter of the two-fold-symmetric
vortex structure. Recent numerical calculations
support the scenario that a structure phase transition for the vortex
profile convolute with the uniform $\top$-breaking phase transition.~\cite{Li}
Such an exotic vortex can be possibly observed either by magneto-optical
spectroscopy, or more possibly by its response to a `rotating'
applied current ${\bf J}=J(\cos\omega t\hat{x}+\sin\omega t\hat{y})$.
At low fields and in a high-quality crystal,
there should be a resonance in the vortex dissipation at a frequency
comparable to the scale of the energy barrier for the vortex to rotate
by an angle of $\pi/2$ or $\pi$.

The mixing of the two order parameters $S$ and $D$ gives a nontrivial
upper critical field, $B_{c2}$.
In our case, the linearized GL equations are:
\begin{eqnarray*}
2\alpha_S S+K[2\Pi^2S+(\Pi_x^2-\Pi_y^2)D]=ES;\\
\alpha_DD+K[\Pi^2D+(\Pi_x^2-\Pi_y^2)S]=ED.
\end{eqnarray*}
where we have add an eigen energy term $E$ on the right hand side of the 
equations. $E=0$ corresponds to the solution of the GL equations.
The condition that the ground state energy $E_g=0$ determines the upper
critical field. This problem has been treated in Refs.~\cite{Berlinsky,Franz}
 where an implicit solution for $B_{c2}$ was obtained.
Here we give an explicit solution along the line of Sigrist {\it et al}
.~\cite{Sigrist} By the inspection that $[\Pi_x,\Pi_y]=2ieB$, 
we can define the bosonic operators $a$ and $a^{\dagger}$ such
that $[a,a^{\dagger}]=1$, with
\begin{eqnarray*}
a=(\Pi_x+i\Pi_y)/\sqrt{4eB}; \ \ a^{\dagger}=(\Pi_x-i\Pi_y)/\sqrt{4eB}.
\end{eqnarray*}
Here $B$ is the magnetic induction, the fluctuation of which at
the upper critical field can be neglected.
In terms of the bosonic operators, the above linearized GL equations becomes
\begin{eqnarray*}
& &[4KeB(2\hat{n}+1)+2\alpha_S]S\\
& &+2KeB(aa+a^{\dagger}a^{\dagger})D=ES,\\
& &2KeB(aa+a^{\dagger}a^{\dagger})S\\
& &+[2KeB(2\hat{n}+1)+\alpha_D]D=ED.
\end{eqnarray*}
Here $\hat{n}=a^{\dagger}a$ is the Landau quantum number. $S$ and $D$
can be expanded in terms of the Landau levels $|n\rangle$ as
$(S,D)=\sum_n (a_n,b_n) |n\rangle$. If there were
no coupling between $S$ and $D$, we would obtain the usual result that
the lowest Landau level $n=0$ determines the upper critical field.
In our case, however, the eigenvalue problem involves coupling between
all next-nearest Landau levels. Expanding the order parameters up to
$n=2$, we can find the ground state energy $E_g$ variationally.
The upper critical field is obtained by setting $E_g=0$, and is
$B_{c2}=[5-\alpha_{SD}+\sqrt{\alpha_{SD}^2+6\alpha_{SD}+25}]B_0/8$
with $B_0=-\alpha_D/2eK$ being the temperature dependent bare upper
critical field for the pure $d_{x^2-y^2}$-
superconductors. In full dimensions, $B_0=-\alpha_D\hbar c/2eK=\Phi_0/2\pi\xi^2$ with $\Phi_0=2\pi\hbar c/2e=hc/2e$ and
$\xi^2=-K/\alpha_D$. It should be pointed out that the above variational
ground state is closely similar to that obtained in Refs.\cite{Berlinsky,Franz},
where $S$ are given by a difference between two
gaussian functions. This is clear from the fact that $S$ 
has two nodes in such a case, as the Landau state with $n=2$ does. Indeed, these solutions
behave similarly: $B_{c2}\rightarrow B_0$ at $T\rightarrow T_D$
(or $\alpha_{SD}\rightarrow \infty$), but
it is always larger than $B_0$, developing an upward curvature in the
temperature dependence near and below $T_D$.~\cite{Berlinsky,Franz} (It diverges in the unphysical region $\alpha\rightarrow -\infty$). This is due to the very effect that
the sub-dominant channel contributes excess energy lowering by adjusting
the shape of the vortex.

Finally, the Abrikosov vortex lattice has also been addressed in the
literature.~\cite{Ren-Xu,Berlinsky,Franz,Li} 
The nontrivial four-fold or two-fold vortex structures have important impact
on the lattice structure. Due to the intrinsic anisotropies
arising from $|S|^2$ they favor
generally an oblique vortex lattice.~\cite{Ren-Xu,Berlinsky,Franz,Li}
Of course, with increasing temperatures,
the amplitude of $S$ decreases, and one generally expects a crossover to
the hexagonal vortex lattice near $T_D$.
Another way of seeing this is as follows.
Since the vortex lattice can be constructed in terms of the 
(highly degenerated) ground state
wave functions for the linearized GL equations,
it can {\it not} be strictly hexagonal if the amplitude
of the $n=2$ Landau level is finite.~\cite{Sigrist} Franz {\it et al}
were also be able to conclude that the orientational angle of the
vortex lattice with respect to the crystal axis, $\alpha$, is no longer
arbitrary (as in the case of a single order parameter), but can only take
the four possible values $\alpha=\pm \pi/2,\pm\pi$,~\cite{Franz}
as was found in numerical simulations.~\cite{Wang}
In the limit of $\alpha_{SD}\rightarrow\infty$ (or $T\rightarrow T_D$),
however, the amplitude for $|n=2\rangle$ vanishes and
we recover the usual result for a pure $D$ order parameter.

Before closing this section, we would like to point out that all conclusions
drawn for $d_{x^2-y^2}+is$ superconductors are also true for
the $d_{xy}+is$ superconductors due to the apparent symmetry.

\section{ $d_{x^2-y^2}+i d_{xy}$ superconductors}
We consider the case $T_{D}>T_{D'}$.
Similar to the above case, there is also a second order $\top$-breaking
phase transition. Here the transition temperature $T_{DD'}$ is given by
$\alpha_{D'D}\equiv \alpha_{D'}/|\alpha_D|<-1/3$ (with $\alpha_D<0$),
or $T_{D'D}=\sqrt{T_{D'}^3/T_D}$. Below $T_{DD'}$ and in the uniform phase,
$D'=\pm i \eta D$ with $\eta=\sqrt{-(1+3\alpha_{D'D})/(3+\alpha_{D'D})}$.
However, above $T_{DD'}$ no $D'$-component can be induced by a vortex
in the absence of a mix-gradient term coupling $D'$ with $D$.
Below $T_{DD'}$, symmetry requires that both components wind in the same
manner, e.g., $D,D'\sim e^{i\theta}$.
Also because of this, the upper critical field is independent of
$\alpha_{D'D}$ so long as $\alpha_{D'D}>-1$ (or simply $T>0$). In this
case, $D'/D=0$ at the upper critical field, even though $D'\neq 0$ in the
absence of the field.

In fact, because $D$ and $D'$ are not coupled
by gradients, the system has two meta-stable states with $D'=0$
and $D'\neq 0$ respectively. The stable state is that with
lower free energy. This is related to the fact that the $\top$-breaking
phase is frustrated at the vortex core if $\alpha_{D'D}>-1$:
While winding simultaneously for $D$ and $D'$
increases the kinetic energy of the
superfluid, the nonzero $D'$-component may lower the homogeneous
energy at low fields. We believe the competing energies may
drive a transition from the
$d_{x^2-y^2}+id_{xy}$-wave vortex state to a pure $d_{x^2-y^2}$-wave
vortex state at a field lower than the upper critical field.
By the observation that the system is in the
$d_{x^2-y^2}+id_{xy}$-wave phase at zero field but is {not} at the upper
critical field, and that the system would be in the
$d_{x^2-y^2}+id_{xy}$-wave vortex state at all fields when $\alpha_{D'D}=-1$,
there must be at least one {\it field-induced} phase transition
below $T_{DD'}$. The transition field is zero and $B_{c2}$
for $T=T_{DD'}$ and $T=0$, respectively, and should generally increase
with decreasing temperature. Above the transition line is the region of
pure $d_{x^2-y^2}$ mixed state.
Such a transition would predict that the quasi-particle excitation gap (
which should be proportional to $|D'|$)
decreases with increasing field below the transition line, and eventually
vanishes above the transition line, where residual density of states
can arise from the Doppler-shift due to the supercurrent around the vortex.
~\cite{Aubin,Kubert}
(The field-induced low-lying quasiparticle states are possible only for a nodal pairing
function).
This translates that at $T<T_{DD'}$ the thermal conductivity should increase
rather than decrease with increasing field, being consistent with the general
trend of the data at sub-Kelvin temperatures
in Ref.~\cite{Aubin}. But the concrete field dependence of the thermal
conductivity may be different. In fact, the mechanism proposed in
Ref.~\cite{Aubin} rules out  the $D'$-component.

\section{Strong coupling effects: phenomenological theory}
As the electrons in unconventional superconductors
are strongly correlated because of their $d$-shell character, 
we now incorporate the strong coupling effects in a pure phenomenological
manner by including the Zeeman energy.
This energy follows from the interaction between the magnetic
field and the angular momentum of the Cooper pair via strong coupling effects.
The underlying mechanism is simple: Cooper pairs with a definite projection
of internal orbital momentum will respond to the direction of the vortex
supercurrent, which is determined by the magnetic field.~\cite{Tokuyasu}
The desired energy density is~\cite{Sigrist}
\begin{eqnarray}
& &-4\pi e h\int d^2{\bf R}g(R)\Delta_{\bf R}^*{\bf R}\times
\frac{\nabla_{\bf R}}{i}\Delta_{\bf R}\nonumber\\
& &\propto ih(D^*D'-{\rm c.c.}), \label{Strong}
\end{eqnarray}
where  $h=\nabla\times A$ is the local magnetic induction,
 $g(R)$ is representative of the strong coupling effect, and
$\Delta_{\bf R}$ is Cooper-pair wave function in the
center-of-mass frame, i.e., the inverse Fourier transform of
$\Delta_k=S+D\cos 2\theta_k+D'\sin 2\theta_k$ subject to the weak
coupling condition $|k|=k_F$:
\begin{eqnarray*}
& &\Delta_{\bf R}\sim \int \Delta_k e^{i{\bf k\cdot R}} d\theta_k\\
& &=SJ_0(kR)+(D\cos 2\theta_R+D'\sin 2\theta_R)J_2(kR),
\end{eqnarray*}
with $J_n(kR)=\int\cos n\phi e^{ikR\cos\phi}d\phi$, and $\theta_R$ being
the angle between ${\bf R}$ and the $a$-axis in the copper-oxide plane.
Here we have adopted the local limit by neglecting the spatial variation
of the order parameters in evaluating the strong coupling term, since
$\lambda_F\ll \xi$ with $\xi$ being the coherence length of the superconductor.
Note that the final result of the strong coupling term does not include
the $S$ component. This is not accidental but results from symmetry.
For convenience we rewrite the {\it non-diamagnetic} contribution (\ref{Strong})
in the form as $\delta k K (2ieh)(D^*D'-{\rm c.c})$
with the only phenomenological dimensionless parameter $\delta k$.
Notice that the Zeeman term can also be rewritten as gradient terms
$\delta k K[(\Pi_xD)^*\Pi_yD'-(\Pi_yD)^*\Pi_xD'+{\rm c.c.}]$
due to the fact that $[\Pi_x,\Pi_y]=2ieh$.

To simplify the matter, here we only consider the $d_{x^2-y^2}+id_{xy}$
superconductor in the presence of strong coupling effect.
In this case, the new free energy reads,
\begin{eqnarray}
F=& &\frac{N(0)}{2}\int_{\Omega}\alpha_D|D|^2+\alpha_{D'}|D'|^2\nonumber\\
& &+\gamma\left[\frac{3}{8}|D|^4+\frac{3}{8}|D'|^4\right.\nonumber\\
& &+\left.\frac{1}{4}|D|^2|D'|^2+\frac{1}{8}(D^*D'+{\rm c.c.})^2\right]\nonumber\\
& &+K\left[|\Pi D|^2+|\Pi D'|^2+2ieh\delta k(D^*D'-{\rm c.c.})\right]\nonumber\\
& &+\int_{\Omega} (\nabla\times A)^2/8\pi \label{Eq:NFenergy}.
\end{eqnarray}

In the case of $\alpha_D\equiv\alpha_{D'}<0$ and $\delta k\neq 0$,
it is known that there could be a first-order phase
transition from the Meissner state to a state with a sudden penetration of
a {\it finite} density of vortices at a lower critical field
$H_{c1}$.\cite{pwave} (The transition is of first
order since $B_{c1}\neq 0$ at $H=H_{c1}$). However, in the extreme
London limit we shall adopt (suitable for high-$T_c$ superconductors),
we expect that both $H_{c1}$ and $B_{c1}$ would be too small to alter our
subsequent discussion, and will be neglected.

Let us now discuss the outcome of Eq.~(\ref{Eq:NFenergy}).
To be concrete but without loss of generality,
we assume $T_c=T_{D}=100K$ and $T_{D'}=10K$.
As discussed in the previous section, at zero magnetic field,
there is a second order $\top$-breaking
phase transition at a critical temperature $T_{DD'}=\sqrt{T_{D'}^3/T_D}\approx 3.16K$.
Below $T_{DD'}$, $D'=\pm i \eta D$. Explicitly,
$|D|^2=(3+\alpha_{D'D})D_0^2/2$ and $|D'|^2=-(1+3\alpha_{D'D})D_0^2/2$.
Here $D_0^2=-\alpha_D/\gamma$. Henceforth we shall normalize
the order parameters in units of $D_0$, and the induction field
by $B_0$ defined previously. 
They are understood as dimensionless quantities unless
specified otherwise. Practically for high-$T_c$ superconductors,
both $D_0$ and $B_0$ saturate at low temperatures (e.g., $T\leq 20K$):
$D_0\sim 2T_c\sim 200K$ and $B_0\sim B_{c2}(0)\sim 40T$--$100T$.
\cite{BCS-scaling}

Following from Eq.~(\ref{Eq:NFenergy}),
in the presence of the non-diamagnetic coupling between the order
parameters and the magnetic field ($\delta k\neq 0$),
a pure $d_{x^2-y^2}$ state is stable only in zero field
above $T_{DD'}$, while a $\top$-broken $d_{x^2-y^2}\pm id_{xy}$ state
is established at all temperatures at finite fields,
albeit with varying amplitude of $D'$. We now concentrate on the
low field regime and in the London limit, so that the kinetic energy
of the superfluid and the inhomogeneity due to vortices
can be safely neglected. The same methodology has been applied in
Ref.~\cite{Laughlin}. At $T>T_{DD'}$ and low fields, the order parameters
can be treated perturbatively, and are, up to the second order
in $\epsilon\equiv\delta kB$ ($>0$ for ease of presentation) with $B$
as the spatial average of the local induction $h$,
\begin{eqnarray}
|D|&=&2/\sqrt{3}-\epsilon |D'|/4,\label{Eq:D}\\
|D'|&=&(\epsilon/\sqrt{3})/(\alpha_{D'D}+1/3).\label{Eq:D'}
\end{eqnarray}
This amounts to a minimum gap on the Fermi surface
\begin{eqnarray}
& &\Delta_0\approx 2|D'|T_c\sim 2\epsilon T_DT_{D'D}/(T-T_{D'D})\nonumber\\
& &\propto \delta kB/(T-T_{D'D})
\label{Eq:GapMin}
\end{eqnarray}
at $\epsilon\ll 1$ and $T>T_{D'D}$ (not very close to $T_{D'D}$).
Neglecting for the moment the suppression due to
quasi-particle scattering at the vortices, we observe that
the thermal conductivity $\kappa_e$ at $T>T_{D'D}$ in the presence of
a minimum gap $\Delta_0$ is given by,
\begin{eqnarray}
& &\kappa_e(T,B)/\kappa_e(T,0)\sim(1/T)\exp(-\Delta_0/T)\nonumber\\
& &\sim (1/T)\exp[-{\rm const.}\times \delta k B/(T-T_{D'D})T].
\label{Eq:kappa}
\end{eqnarray}
It should be appreciated that this exponential decay in $B$ is increasingly
sharpened by $T(T-T_{D'D})$ instead of only $T$.
In Eq.~(\ref{Eq:kappa}) a power law $\kappa_e(T,0)\propto T^2$ is used,
which follows from the
fact that above $T_{D'D}$ the zero-field state is a pure $d_{x^2-y^2}$ state.
Further we have implicitly neglected the change in
$|D|$, which is merely of the second order in $\epsilon$ from Eq.~(\ref{Eq:D}).
The exponential decay develops a crossover at $\Delta_0\sim T$, or,
\begin{equation}
B_k\propto T(T-T_{D'D}),\label{Eq:crossover}
\end{equation}
which indeed resembles a scaling law $B_k\propto T^2$
addressed by Krishana {\it et al}.~\cite{Krishana}
As an example, we plot
$|D'|$ and $\kappa_e(T,B)/\kappa_e(T,0)$ in this context in Figs.~1(a) and (b),
respectively. Here $|D'|$ is calculated exactly from the optimization of
the free energy subject to the neglecting of kinetic energy and field energy,
which should be reliable at low fields. Defining a criterion for the kink
transition (more precisely, cross-over), the parabolic-like contour
lines unambiguously reproduce the scaling law for the kink field.~\cite{Krishana}
The sharpening feature of the crossover is also clear.
Including the additional suppression of $\kappa_e$
from quasi-particle scattering by vortices would
render even better qualitative agreement with the experimental results,
but we shall not go into further details. Interestingly,
Eq.~(\ref{Eq:crossover}) predicts a downward curvature
of the $B_k(T)$ curve in a log-log plot on the low temperature side,
which seems to be the case in Fig.~4(b) of Ref.\cite{Krishana}.
We have also performed exact calculation of
the full set of GL equations ensuing from Eq.(\ref{Eq:NFenergy}) to find
that the results in the low field regime remains essentially unchanged, while
at intermediate fields, $|D'|$ saturates a while and then diminishes gradually
together with $|D|$ near the upper critical field $B_{c2}$.
The latter is given by, in dimensionless form,
$B_{c2}=\left[1-\alpha_{D'D}+\sqrt{(1+\alpha_{D'D})^2-4\alpha_{D'D}\delta k^2}\right]/2(1-\delta k^2)$.
At $B_{c2}$ the ratio $|D'|/|D|$ is given by
$|D'|/|D|=\delta k B_{c2}/(\alpha_{D'D}+B_{c2})$.
This means that $\top$-breaking is retained up to the upper critical field.
However, it is clear, in conjunction with Fig.~1(a),
that the strong coupling effect vanishes
near the critical temperature $T_c$ (where $\alpha_{D'D}\rightarrow\infty$),
a necessary ingredient for the theory to be compatible with the
general properties of high-$T_c$ superconductors at temperatures near $T_c$.
Also interestingly, in the context of this theory, (i) the power law
$\kappa_e\propto T^2$ is invalid at finite fields above $T_{D'D}$,
which seems to be consistent with the experimental data in
Ref.~\cite{Krishana}, and would be violated below $T_{D'D}$ even in
the absence of magnetic field; (ii) $\kappa_e$ increases with increasing
magnetic field below $T_{D'D}$.
On the other hand, the theory does not involve a phase transition
for the development of a plateau in $\kappa_e$,
because there is no further symmetry breaking at finite fields
in our theory.
\begin{figure}
\epsfxsize=10cm
\epsfbox{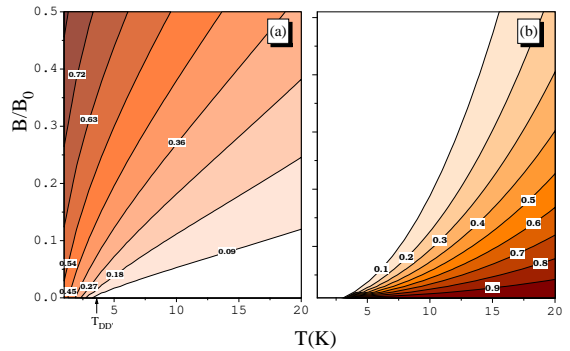}
\caption{Gray-scale contour plots of (a) $|D'|$ and (b) $\kappa_e(T,B)/\kappa_e(T,0)
\approx (1/T)\exp(-\Delta_0/T)$
as functions of $B$ and $T$. Here $\delta k=1/2$ and $\Delta_0=2|D'|T_c$
with $T_c=100K$. $T_{DD'}\sim 3.16K$ is the zero-field
$\top$-breaking transition temperature. See the text for details.}
\end{figure}

\section{Summary}
We have developed a GL theory with respect to the three relevant singlet
pairing channels in unconventional superconductors. The case of $d+is$
superconductors has been discussed. The novel two-fold symmetric
structure of vortices below the $\top$-breaking temperature is predicted.
The case of $d_{x^2-y^2}+id_{xy}$ superconductors has been addressed
in detail. In the absence of strong coupling effects, we find that the
sub-dominant $d_{xy}$-wave order parameter can not be induced by vortices
above the $\top$-breaking temperature, in contrast to the case in a
$d+is$ superconductor. Below the $\top$-breaking temperature, we predict
a novel field-induced first-order phase transition from the
$d_{x^2-y^2}+id_{xy}$ vortex state to the $d_{x^2-y^2}$ vortex state at
higher fields. By including the strong coupling effect 
phenomenologically, the $d_{x^2-y^2}+id_{xy}$ state can be induced by the magnetic
field even above the zero-field $\top$-breaking temperature. Near and
above this transition temperature, the induced $d_{xy}$-wave order parameter
may be compared with some results for the abnormal thermal conductivity
reported in Ref.~\cite{Krishana}. Although our theory could give
the general trend that the thermal conductivity would increase with
increasing field below the $\top$-breaking temperature,
it can not result in a pure $d_{x^2-y^2}$ state near the zero temperature,
which was claimed to be essential to explain the sub-Kelvin
thermal conductivity in Ref.~\cite{Aubin,Kubert}. If that is the case,
we argue that there are two possibilities for the
material used in Ref.~\cite{Aubin}: either some significant change occurs
for the $d_{xy}$-channel near zero temperature such that the $d_{xy}$-wave
order parameter is suppressed drastically or no $d_{xy}$-channel is
present at all.

It should be pointed out that the $s$- and $d_{xy}$- channels are
only two, but not all of  possible sub-dominant singlet pairing channels
in a $d_{x^2-y^2}$-dominant superconductor. We have neglected, {\it e.g.},
the $k_xk_y(k_x^2-k_y^2)$-channel (in the $A_{2g}$ representation),
which would appear in the case of a more general pairing interaction,
or in a higher-order expansion of the gap equation.~\cite{Zhu}
This order parameter could also be induced by spatial inhomogenieties of
the $d_{x^2-y^2}$ order parameter, and would indirectly induce a $d_{xy}$
order parameter. However, odd-parity order parameters can not be induced
in singlet-pairing superconductors.

In this paper, we have only performed accessible analytical as well as
qualitative discussions on the theory. More exact and numerical results
are awaited. The theory is also highly useful for
studying the vortex dynamics of unconventional superconductors.~\cite{Wang}

\acknowledgments{This work was supported by the HKU block grant,
the RGC grant of Hong Kong under No. HKU7116/98P and HKU 7144/99P,
 the National Natural Science Foundation
of China, and in part, by Ke-Li Fellowship financed by Sanzhu Co. Ltd.
in Shandong. ZDW ackknowledges the hospitality of the National Center
for Theoretical Sciences at Taiwan, where the final version was completed.}
\\
$*$ Corresponding author: zwang@hkucc.hku.hk


\begin{thebibliography}{4}
\bibitem{Tsui} C. C. Tsuei, {\it et al}, Phys. Rev. Lett. {\bf 73},
593(1997); J. R. Kirtley, {\it et al}, Nature {\bf 373}, 225(1995).
\bibitem{Covington-Fogelstrom} M. Covington, {\it et al}, Phys. Rev.
Lett. {\bf 79}, 277(1997); M. Fogelstr\"{o}m, D. Rainer, and J. A. Sauls,
{\it ibid} {\bf 79}, 281(1997).
\bibitem{Kirtly} J. R. Kirtley, {\it et al}, Phys. Rev. B {\bf 51},
12057(1995). For a theoretical interpretation, see, e.g.,
M. Sigrist, D. B. Bailey, and R. B. Laughlin, Phys. Rev. Lett.
{\bf 74}, 3249(19945).
\bibitem{Krishana} K. Krishana, {\it et al}, Science {\bf 277}, 83(1997).
\bibitem{Aubin} H. Aubin, K. Behnia, S. Ooi, and T. Tamegai, Phys. Rev.
Lett. 82, 624(1999).
\bibitem{Laughlin} R. B. Laughlin, Phys. Rev. Lett. {\bf 80}, 5188(1998).
\bibitem{Ghosh} H. Ghosh, Europhys. Lett. {\bf 43}, 707(1998);
Phys. Rev. B {\bf 59}, 3357(1999); Cond-mat/9904375.
\bibitem{Joynt} K. A. Musaelian, J. Betouras, A. V. Chubukov, and
R. Joynt, Phys. Rev. B {\bf 53}, 3598(1996); M. Liu, D. Y. Xing, and Z. D. Wang,
Phys. Rev. B {\bf 55}, 3181(1997).
\bibitem{Note} It is understood that the temperature dependencies
in the parameters of the free energy functional is valid strictly near
the superconducting temperature. At low temperatures, $\gamma$
and $K$ can be shown to be roughly temperature independent from
more exact analysis. On the other hand, $\alpha_i$ is better approximated
as $\alpha_i\sim (T-T_i)/T_i$ at low temperatures in order to
reproduce the fact that $\Delta_k$ saturates as
$T\rightarrow 0$. However, these complications are not crucial for our
main conclusions in the text.
\bibitem{Ren-Xu}Y. Ren, J. H. Xu, and C. S. Ting, Phys. Rev. Lett. {\bf 74},
3680(1995).
\bibitem{Berlinsky} A. J. Berlinsky, A. L. Fetter, M. Franz, C. Kallin, and P. I. Soininen, Phys. Rev. Lett. {\bf 75}, 2200(1995).
\bibitem{Franz} M. Franz, C. Kallin, P. I. Soininen, A. J. Berlinsky, and A.
L. Fetter, Phys. Rev. Lett. {\bf 53}, 5795(1996).
\bibitem{Li} Q. Li, Z. D. Wang, and Q.-H. Wang, Phys. Rev. B {\bf 59}
, 613 (1999); {\it ibid} {\bf 60}, (in press, 1999). 
\bibitem{Volovik} G. E. Volovik, Pis'ma Zh. Eksp. Teor. Fiz. {\bf 58}, 457(1993), JETP Lett. {\bf 58}, 469(1993).
\bibitem{Wang}Q. Wang and Z. D. Wang, Phys. Rev. B
{\bf 54}, R15645(1996); Z. D. Wang and Q. H. Wang,
{\it ibid}, {\bf 55}, 11756(1997).
\bibitem{Sigrist} M. Sigrist, and K. Ueda, Rev. Mod. Phys. {\bf 63},
239(1991).
\bibitem{Kubert} C. K\"{u}bert, and P. Hirschfeld, Phys. Rev. Lett. 80,
4963(1998).
\bibitem{Tokuyasu} T. A. Tokuyasu, D. W. Hess, and J. A. Sauls,
Phys. Rev. B 41, 8891 (1990).
\bibitem{pwave} Z. D. Wang, and Q. H. Wang, Phys. Rev. B {\bf 57}, R724(1998);
Q. H. Wang, and Z. D. Wang, {\it ibid} {\bf 57}, 10307(1998).
\bibitem{BCS-scaling} C. Panagopoulos and T. Xiang, Phys. Rev. Lett. {\bf 81}, 2336(1998).
\bibitem{Zhu} Wonkee Kim, Jian-Xin Zhu, and C. S. Ting, Phys. Rev. B
{\bf 57}, 13403 (1998).
\end{thebibliography}
\end{document}